\documentclass[secnumarabic,amssymb, amsmath,twocolumn, 
nofootinbib,tightenlines, nobibnotes, aps, prl]{revtex4}
\usepackage{graphicx}
\begin{document}
\newcommand{\be}{\begin{equation}}
\newcommand{\ee}{\end{equation}}
\newcommand{\ba}{\begin{eqnarray}}
\newcommand{\ea}{\end{eqnarray}}
\newcommand{\ts}{\textstyle}

\bigskip
\vspace{2cm}
\title{Long-distance radiative corrections to the di-pion tau lepton 
decay}
\vskip 6ex
\author{F. Flores-B\'aez}
\email{pacob@fis.cinvestav.mx}
\author{A. Flores-Tlalpa}
\email{afflores@fis.cinvestav.mx},
\author{G. L\'opez Castro}
\email{glopez@fis.cinvestav.mx}
\affiliation{Departamento de F\'{\i}sica, Cinvestav,  
Apartado Postal 14-740, 07000 M\'exico D.F., M\'exico}
\author{G. Toledo S\'anchez}
\email{toledo@fisica.unam.mx}
\affiliation{Instituto de F\'{\i}sica, UNAM, A.P. 20-364, 01000 M\'exico 
D.F, M\'exico}
\bigskip

\bigskip

\bigskip

\begin{abstract} 
 We evaluate the model-dependent piece of $O(\alpha)$ long-distance 
radiative corrections to $\tau^- \to  \pi^- \pi^0\nu_{\tau}$ decays by
using a meson dominance model. We find that these corrections to the 
di-pion invariant mass spectrum are smaller than in previous calculations 
based on chiral perturbation theory. The corresponding correction to the 
photon inclusive rate is tiny ($-0.15\%$) but it can be of relevance 
when new measurements reach better precision.

\end{abstract}

\maketitle
\bigskip

\bigskip

\section{1. Introduction}

 The decay  $\tau^{\pm} \to \pi^{\pm}\pi^0\nu_{\tau}$ ($\tau_{2\pi}$) is 
the dominant mode of $\tau$ lepton decays. Observables associated to this 
decay channel have been measured with great precision. The world average 
value of the $\tau_{2\pi}$ branching fraction has attained an accuracy of 
0.5\% \cite{pdg}. Similarly, the weak pion form factor has been 
measured with high  precision, providing a  valuable input to compute 
the dominant hadronic contribution to the muon anomalous magnetic moment 
\cite{Davier:2005xq}. Further improvements in measurements of such 
observables are expected at B factories, where BaBar and Belle 
collaborations have 
recorded already about $10^9$ tau lepton decays \cite{Igonkina:2006tv}. 
Therefore, present and future precision measurements of  these observables 
demand the knowledge of radiative corrections for a correct comparison of 
theory and experiment. 

 On another hand, measurements of $\tau_{2\pi}$ decays provide a clean 
environment to test the Conserved Vector Current (CVC) hypothesis.    
As is well known, the two-pion production in tau decays and $e^+e^-$ 
annihilations is driven by the isovector piece of the vector current:  
$\bar{q}\gamma_{\mu}\tau_i q/2$, where $\bar{q}=(\bar{u},\bar{d})$. Thus, 
measurements of the pion form factor in these two reactions and of the 
$\tau_{2\pi}$ branching ratio can be used to provide one of the most 
precise tests of the CVC hypothesis.   In order to perform a test of  the 
CVC hypothesis at a few of per mille level,  the isospin breaking effects 
must be included appropriately  \cite{Ghozzi:2003yn}. The two identifiable 
sources  of isospin breaking corrections in relating the $\tau_{2\pi}$ 
and $e^+e^- \to \pi^+\pi^-$ reactions are the mass  difference of charged 
and neutral pions and, again, the radiative  corrections. 

 The dominant piece of short-distance electroweak radiative corrections 
were 
computed long ago \cite{Sirlin:1977sv}, and 
improvements that include resummation of dominant logs, subleading 
electroweak corrections \cite{Braaten} and resummation of strong 
interactions corrections were subsequently incorporated \cite{Erler}. The 
effects of long-distance corrections to the hadronic spectrum of 
$\tau_{2\pi}$ decays were computed only recently in Refs. 
\cite{Cirigliano:2001er, Cirigliano:2002pv} in the framework of Chiral 
Perturbation Theory supplemented with anomalous terms for the axial 
couplings \cite{WessZumino}. Since $\tau$ decays involve momentum  
transfers far from the chiral limit, an independent calculation of 
radiative corrections based on different model considerations is 
important. 
  In this paper we evaluate the model-dependent contributions to 
long-distance radiative corrections in $\tau_{2\pi}$ decays based on a 
meson dominance model that was used in ref. \cite{Flores-Tlalpa:2005fz} to 
study the corresponding radiative tau decays ($\tau \to 
\pi\pi\nu\gamma$). We focus here on the long-distance radiative 
corrections of $O(\alpha)$ to the  di-pion spectrum and to the decay rate 
of  $\tau_{2\pi}$ decays.

\section{2. Long-distance corrections to the di-pion spectrum}

The hadronic spectrum of two-pions in tau decays, corrected by $O(\alpha)$ 
radiative corrections, is built out of three contributions:  
\be
\frac{d\Gamma(\tau_{2\pi(\gamma)})}{dt} = \frac{d\Gamma^0}{dt} 
+\frac{d\Gamma^1_v}{dt} + \frac{d\Gamma^1_r}{dt} \ .
\ee
The superindexes in the r.h.s. terms denote the order in $\alpha$ and the 
subindex $v (r)$ refers to the virtual (real) corrections, while $t$ is 
the square of the two-pion invariant mass. When we deal with the photon 
inclusive spectrum, integration over all energies of real photons should 
be  done in the last term.

After regrouping the effects of all radiative corrections, the corrected 
spectrum can be rewritten as \cite{Cirigliano:2001er, Cirigliano:2002pv}:
\ba
\frac{d\Gamma(\tau_{2\pi(\gamma)})}{dt} &=& \frac{G_F^2m_{\tau}^3S_{EW}
|V_{ud}|^2}{384\pi^3}\beta_{\pi^+}^3\left(1-\frac{t}{m_{\tau}^2} \right)^2
\nonumber \\
&& \ \ \times \left(1+\frac{2t}{m_{\tau}^2} \right) |f_+(t)|^2 G_{EM}(t)\ 
,
\ea
where $G_F$ denotes the Fermi constant, $|V_{ud}|=0.9740$ is the 
Cabibbo-Kobayashi-Maskawa $ud$ matrix element, and $\beta_{\pi^+}$
is the pion velocity in the di-pion rest frame. 

 The weak form factor of  the pion parametrizes the hadronic amplitude (an 
additional scalar form factor that is allowed by Lorentz covariance 
induces negligible small isospin breaking effects in eq. (2)): 
\be
\langle \pi^+(q)\pi^0(q')|\bar{u}\gamma_{\mu}d|0\rangle = \sqrt{2}  
f_+(t)(q-q')_{\mu} \ .
\ee 
Our numerical results in this paper were obtained using the pion form 
factor described in ref. \cite{Flores-Tlalpa:2005fz}. If we use the 
form factor of ref. \cite{Guerrero:1997ku} we obtain almost identical 
numerical results.

The effects of short-distance corrections to the hadronic decay rate are 
encoded in $S_{EW}=1.0236 \pm 0.0003$ \cite{Erler}, which includes the 
resummation of dominant logs, subleading corrections \cite{Braaten} and 
resummation of strong interactions effects \cite{Erler}. The 
factor 
$G_{EM}(t)$  introduced in eq. (2) contains  all the effects of 
long-distance 
electromagnetic corrections of $O(\alpha)$ and is the main focus of this 
paper. In terms of the rates  appearing in eq. (1) it is defined as 
follows:
\be
G_{EM}(t) =1 + \frac{\ts \frac{\ts d\Gamma^1_v}{\ts dt} + 
\frac{\ts d\Gamma^1_r}{\ts dt}}{\ts \frac{\ts d\Gamma^0}{\ts dt}}\ .
\ee
This function is finite in the limit of infrared photons since the 
divergent terms in virtual and real corrections cancel each other. 
$G_{EM}(t)$ depends on the details of the model used to describe the 
interactions of photons with the hadrons. A useful theorem due to 
Burnett and Kroll and to Zakharov, Kondratyuk and Ponomarev  
\cite{Burnett:1967km}, allows us 
to split the di-pion spectrum of radiative $\tau_{2\pi}$ decays into its 
model-independent and model-dependent parts:
\be
\frac{d\Gamma^1_r}{dt}=\frac{d\Gamma^1_r(m.i.)}{dt}+ 
\frac{d\Gamma^1_r(m.d.)}{dt}\ .
\ee
The first term contains the infrared divergence in the photon energy that 
is necessary to cancel the corresponding divergent term in the 
virtual correction. It is model-independent in the sense that only the 
hadronic structure already present in the non-radiative decay contributes. 
This term has its origin in the photons emitted off the charges of 
$\tau^-$ and  $\pi^-$ as if they were taken as point particles.

  The electromagnetic structure of hadrons and other model-dependent 
couplings of photons to hadrons enter into the second term of eq. (5), 
which is regular for infrared photons. In 
ref. \cite{Cirigliano:2002pv} the model-dependent contributions were 
treated in the framework of Chiral Perturbation Theory supplemented with  
chiral anomalous terms. According to their results, the model-dependent 
terms affect the values of $G_{EM}(t)$  by --0.5\% in most of the region 
of $t$. As it was 
recognized in ref. \cite{Cirigliano:2002pv}, one should not expect that 
some of their model-dependent axial contributions remains adequate   
when they are extrapolated to large momenta as encountered in $\tau$ 
decays. On the other hand, 
since the energy released to hadrons in $\tau_{2\pi}$ decay can be as 
large as $m_{\tau}$, some of the well established light resonances can 
be produced as intermediate states on their mass-shell. Given all these 
considerations, in ref. 
\cite{Flores-Tlalpa:2005fz} we have used a meson dominance model 
to study the effects of model-dependent contributions 
in different observables associated with radiative $\tau_{2\pi}$ decays.
We have found that some of these observables exhibit important 
differences with respect to the 
results of ref. \cite{Cirigliano:2002pv}. These differences arise from  
the model-dependent terms and  are due essentially to the production and 
decay of the  $\omega(782)$ vector meson as an intermediate state in 
radiative tau decay. In the following we explore the consequences of such 
meson dominance model contributions to the model-dependent part of  
long-distance radiative corrections.

  Using the decomposition shown in eq. (5), we can split the 
radiator function $G_{EM}(t)$ as it was proposed in ref. 
\cite{Cirigliano:2002pv}:
\be
G_{EM}(t)=G_{EM}^0(t)+G_{EM}^{m.d.}(t)\ ,
\ee
where the first term denotes the model-independent piece 
(tree-level, virtual and real), and the second term includes the 
model-dependent terms. In this paper we use the expression of 
$G_{EM}^0(t)$ calculated in ref. \cite{Cirigliano:2002pv}. We have checked 
that the piece in this term coming from model-independent bremsstrahlung 
is correct, and we 
have focused on the evaluation of the second term of eq. (6) based on the 
model discussed in our previous work \cite{Flores-Tlalpa:2005fz}. 

   In Figure 1 we have plotted 
our values for $G_{EM}(t)$ (solid line). In the same plot we  
also display the values of $G_{EM}^0(t)$ (short-dashed line) and 
the total $G_{EM}(t)$ (long-dashed line) obtained in ref. 
\cite{Cirigliano:2002pv}. The sharp increase of $G_{EM}(t)$ 
close to threshold is due to the phase space suppression of the 
tree-level rate at those energies, the denominator of second term in eq. 
(4).  
This means that corrections can not be fully reliable for values of $t$ 
very close to threshold. We  observe  that our model-dependent 
term differs from that obtained in ref. \cite{Cirigliano:2002pv} for low 
and intermediate values of $t$, finding 
the largest difference (of order 1\%) for $t \approx 1.5$ GeV$^2$. This 
discrepancy in model-dependent terms arise almost completely from the 
contribution of the $\omega(782)$ intermediate state in radiative 
$\tau_{2\pi}$ decays (Figure 2.g, of ref. \cite{Flores-Tlalpa:2005fz}). 
To confirm this, we plot in 
Figure 2 our  results for $G_{EM}(t)$ obtained when we 
exclude the  diagram containing the  $\omega(782)$ vector meson (dotted 
line). It is interesting to see that when we 
exclude the contribution due to the $\omega(782)$ meson, our result 
coincides with the one obtained in ref. \cite{Cirigliano:2002pv} 
(long-dashed line in Figure 2). Equivalently, if we exclude only the 
contributions involving an intermediate  $a_1(1260)$ state (Figures 2.e, 
2.f, 2.j and 2.k of ref.  \cite{Flores-Tlalpa:2005fz}), the values 
obtained for $G_{EM}(t)$ almost overlap the solid line in Figure 1.

\begin{figure}
  \includegraphics[width=6cm, angle=270]{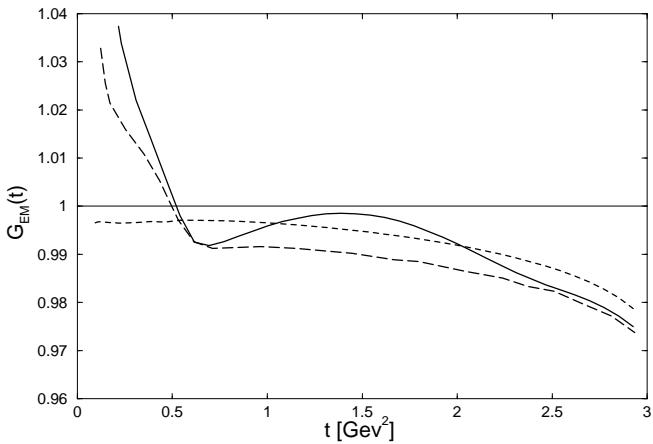}\\
  \caption{Long-distance electromagnetic correction to the 
di-pion spectrum: this paper including all contributions 
(solid-line), model-independent (short-dashed-line) and full 
contributions of ref. \cite{Cirigliano:2002pv} (long-dashed 
line).}\label{fig1} \end{figure}

\begin{figure}
  \includegraphics[width=5.5cm, angle=270]{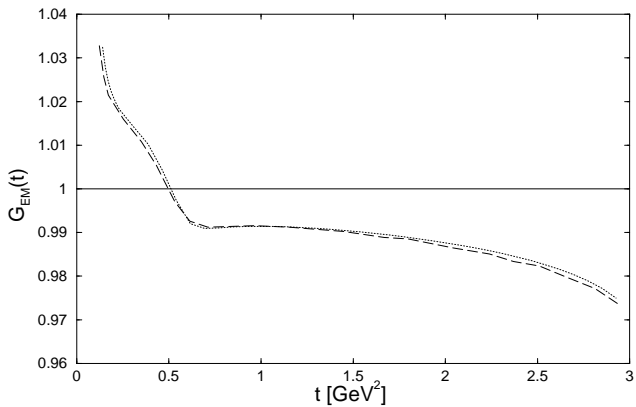}\\
  \caption{Comparison of our results for $G_{EM}(t)$ obtained when we 
exclude the  $\omega(782)$ meson (dotted) and results of ref. 
\cite{Cirigliano:2002pv} (long-dashed).}\label{fig2} 
\end{figure}

   The simple formula ($x\equiv t/m_{\tau}^2$):
\ba
G_{EM}(t)&=& 1.107-1.326x+5.667x^2-10.95x^3\nonumber \\
&& \ \ +9.735x^4-3.276x^5
\ea
provides a good description of $G_{EM}(t)$ in the 
interval $0.18\ \mbox{\rm GeV}^2 \leq t \leq 2.93\ \mbox{\rm GeV}^2$.
It differs numerically from the exact calculation by less than 0.3\% for 
the interval under consideration. It can be useful to 
apply the radiative corrections to future data and/or theoretical 
predictions to the di-pion spectrum of photon inclusive $\tau_{2\pi}$ 
decays.

\section{3. Long-distance corrections to the decay rate}

 Radiative corrections to the decay rate of $\tau_{2\pi}$ are of interest 
for a precise comparison of theoretical and experimental branching ratios 
and also to provide a test of the CVC hypothesis at the few per mille 
level. This 
observable can be obtained by integrating eq. (1) over $t$. As usual, we 
define the radiative corrected decay rate as follows:
\be
\Gamma(\tau_{2\pi(\gamma)})= \Gamma(\tau_{2\pi})\cdot (1+\delta)\ , 
\ee 
where $\delta$ encodes the effects of long-distance corrections, while 
$\Gamma(\tau_{2\pi})$ corresponds to the integrated rate obtained when 
we set $G_{EM}(t)=1$ in eq. (2). 

  If future measurements are able to discriminate real photons of 
energy larger than $\omega_0$, it is convenient to compute the 
radiative corrections $\delta(\omega < \omega_0)$ such that the corrected 
rate include photons  of energies below this threshold. In Table I we show 
the long-distance corrections corresponding to different values of 
$\omega_0$.

\begin{table}[t]
\begin{tabular}{|c|c|c|}
\hline 
$\omega_0$ (MeV) & $\delta(\omega < \omega_0)$ in \% & $\delta(\omega < 
\omega_0)$ in \%\\
 & This work & Ref. \cite{Cirigliano:2002pv} \\
\hline
300 & --0.31& --\\
400 & --0.27 & --\\
500 & --0.23& --\\
600 & --0.19& --\\
700 & --0.16& --\\
800 & --0.15& -- \\
$\omega_{max}$ & -0.15 & -0.38 \\  
\hline
\end{tabular}
\caption{ Long-distance radiative correction to the 
$\tau_{2\pi}$ decay  rate when photons below the threshold energy 
$\omega_0$ are included.}
\end{table}

The long-distance correction corresponding to the {\it photon  inclusive 
rate} is shown in the last row of Table I. Just for comparison, we have 
evaluated this correction for the model of ref. \cite{Cirigliano:2002pv} 
and displayed in the third column of Table I. The largest contribution to 
$\delta$ comes from the model-independent terms ($-0.32\%$). In our model, 
this correction is partially cancelled by the model-dependent contributions
($+0.17\%$). This is much smaller than the contribution of the 
hard-photon component of radiative corrections ($+0.8\%$) estimated in 
ref. \cite{deTroconiz:2004tr}.

\section{4. CONCLUSIONS}

The long-distance radiative corrections of 
$O(\alpha)$ to $\tau_{2\pi}$ decays in the framework of a vector dominance 
model, turns out to be smaller than the one calculated in ref. 
\cite{Cirigliano:2002pv} using Chiral Perturbation theory. The corrections 
to the di-pion mass distribution are less than 1\% in most of the relevant 
values of the $t$. In particular, the discrepancy between measurements of 
the pion  electromagnetic form factors in $\tau$ lepton decays and 
$e^+e^-$  annihiliations \cite{Davier:2005xq} can not be explained by 
radiative  corrections. The long-distance  radiative 
correction to the photon inclusive decay rate of $\tau_{2\pi}$ 
($-0.15\%$) is below the present experimental uncertainties ($\pm 0.5\%$) 
but it can be of relevance for a test of the CVC hypothesis when improved 
measurements become available. From another perspective, we 
can conclude from this and previous calculations that long-distance 
corrections are well under control at a few per mille level. 

\section{Acknowledgements}
 The authors want to acknowledge financial support from Conacyt 
(M\'exico). 

\section{Note added in proofs}

   Model-dependent radiative corrections affect also the calculation of 
the two-pion vacuum polarization contribution to the muon anomalous 
magnetic moment ($a_{\mu}^{LO,\pi\pi}$) extracted from $\tau$ decay data 
(see for example \cite{Davier:2005xq}). The shift in $a_{\mu}^{LO,\pi\pi}$  
produced by such corrections when compared to the uncorrected result is 
given by \cite{Cirigliano:2002pv}:
\ba
\Delta a_{\mu}^{LO, \pi\pi}\! &=&\! \frac{1}{4\pi^3} 
\int_{4m_{\pi}^2}^{m_{\tau}^2}\!\!\! dt K(t)\left[ 
\frac{K_{\sigma}(t)}{K_{\Gamma}(t)}\frac{d\Gamma_{\pi\pi(\gamma)}}{dt}\right] 
\left(\frac{1}{G_{EM}(t)}-1 \right) \nonumber \\
&=& -3.7 \times 10^{-10}\ .\nonumber
\ea
The kernel function $K(t)$ and the kinematical factors 
$K_{\sigma,\Gamma}(t)$ can be found in reference \cite{Cirigliano:2002pv}. 
Our result, shown in the equation above, is almost 4 times larger that the result 
reported in  \cite{Cirigliano:2002pv}. This difference can be traced back 
to the  our model-dependent contribution involving the intermediate 
$\omega(782)$ vector meson (see discussion in section 2) and it was 
not considered in the calculation of ref. \cite{Cirigliano:2002pv}.


\begin{thebibliography}{40}
\bibitem{pdg}
S.  Eidelman et al Review of Particle Properties, Phys. Lett. {\bf
B592}, 1 (2004).
\bibitem{Davier:2005xq}
 See for example: M.~Davier, A.~Hocker and Z.~Zhang,
  arXiv:hep-ph/0507078.
\bibitem{Igonkina:2006tv}
  O.~Igonkina,
  arXiv:hep-ex/0606009.
\bibitem{Ghozzi:2003yn}
 See for example: S.~Ghozzi and F.~Jegerlehner,
  Phys.\ Lett.\ B {\bf 583}, 222 (2004)
  [arXiv:hep-ph/0310181].
\bibitem{Sirlin:1977sv}
  A.~Sirlin,
  Rev.\ Mod.\ Phys.\  {\bf 50}, 573 (1978)
  [Erratum-ibid.\  {\bf 50}, 905 (1978)].
W.~J.~Marciano and A.~Sirlin,
  Phys.\ Rev.\ Lett.\  {\bf 61}, 1815 (1988);
 A.~Sirlin,  Nucl.\ Phys.\ B {\bf 196}, 83 (1982)
\bibitem{Braaten}
 E.~Braaten and C.~S.~Li,
  Phys.\ Rev.\ D {\bf 42}, 3888 (1990).
\bibitem{Erler}
J.~Erler,  Rev.\ Mex.\ Fis.\  {\bf 50}, 200 (2004)
  [arXiv:hep-ph/0211345].
\bibitem{Cirigliano:2001er}
  V.~Cirigliano, G.~Ecker and H.~Neufeld,
  Phys.\ Lett.\ B {\bf 513}, 361 (2001)
  [arXiv:hep-ph/0104267].
\bibitem{Cirigliano:2002pv}
  V.~Cirigliano, G.~Ecker and H.~Neufeld,
  JHEP {\bf 0208} (2002) 002
  [arXiv:hep-ph/0207310].
\bibitem{WessZumino}
J. Wess and B. Zumino, Phys. Lett. {\bf B37}, 95 (1971).
\bibitem{Flores-Tlalpa:2005fz}
  A.~Flores-Tlalpa, G.~Lopez Castro and G. Toledo S\'anchez,
  Phys.\ Rev.\ D {\bf 72}, 113003 (2005)
  [arXiv:hep-ph/0511315].
\bibitem{Guerrero:1997ku}
  F.~Guerrero and A.~Pich,
  Phys.\ Lett.\ B {\bf 412}, 382 (1997)
  [arXiv:hep-ph/9707347].
\bibitem{Burnett:1967km}
  T.~H.~Burnett and N.~M.~Kroll,
  Phys.\ Rev.\ Lett.\  {\bf 20}, 86 (1968);
V.~I.~Zakharov, L.~A.~Kondratyuk and L.~A.~Ponomarev,
  Yad.\ Fiz.\  {\bf 8}, 783 (1968)
  [Sov.\ J.\ Nucl.\ Phys.\  {\bf 8}, 456 (1969)].
\bibitem{deTroconiz:2004tr}
  J.~F.~de Troconiz and F.~J.~Yndurain,
  Phys.\ Rev.\ D {\bf 71}, 073008 (2005).
  [arXiv:hep-ph/0402285].
\end{thebibliography}
\end{document}